\documentclass[prl,superscriptaddress,twocolumn,preprintnumbers,a4,showpacs]{revtex4}
\usepackage{natbib, amsmath, amssymb, psfrag, graphicx, tabls, dcolumn, bm, color}
\begin{document}
\title{ Understanding of the phase transformation from fullerite to
    amorphous carbon at the microscopic level}
\author{M. Moseler}
   \affiliation{Fraunhofer-Institut f\"ur Werkstoffmechanik IWM, W\"ohlerstr. 11, 79108 Freiburg, Germany}
\author{H. Riedel}
   \affiliation{Fraunhofer-Institut f\"ur Werkstoffmechanik IWM, W\"ohlerstr. 11, 79108 Freiburg, Germany}
\author{P. Gumbsch}
   \affiliation{Fraunhofer-Institut f\"ur Werkstoffmechanik IWM, W\"ohlerstr. 11, 79108 Freiburg, Germany}
\author{J. St\"aring}
   \affiliation{Theoretical Physics, G\"oteborg University/Chalmers, Sweden}
\author{B. Mehlig}
   \affiliation{Theoretical Physics, G\"oteborg University/Chalmers, Sweden}
\date{\today}
\begin{abstract}
%
%
We have studied the shock-induced phase transition from fullerite to a dense amorphous 
carbon phase by 
tight-binding molecular dynamics. For increasing hydrostatic pressures $P$, 
the C$_{60}$-cages are found
to polymerise at $P\!<10\!$~GPa, to break at $P \sim 40$GPa  
and to slowly collapse further at $P\!>\!60$ GPa. 
By contrast, in the presence of additional shear stresses,
the cages are destroyed at much lower pressures
($P\!<\!30$ GPa). We explain this fact in terms of a continuum model,
the snap-through instability of a spherical shell.
Surprisingly, the relaxed high-density structures
display no intermediate-range order.
\end{abstract}
\maketitle
Rapid development in cluster science and  
nanotechnology has fostered the hope
that materials exhibiting novel properties
due to the existence of intermediate-range order (IRO)
can be constructed from clusters~\cite{sat96} or nanotubes~\cite{elliott} as building blocks.
Fullerenes (C$_{60}$ molecules) represent one of the rare species of clusters 
available in macroscopic quantities~\cite{kra90}. It is 
not surprising, therefore, that 
fullerene-assembled materials have been studied
for at least a decade now~\cite{tac94}.

The 
extraordinary elastic properties of the molecular cage of C$_{60}$
have inspired many to speculate about new carbon modifications with 
extraordinary mechanical performance~\cite{ruo91}. 
Indeed, in early experiments  very hard and stiff materials 
were synthesised (stable under
ambient conditions) by compressing 
fullerite under high pressure~\cite{nun91}.
Subsequently, these  phases have
been characterised experimentally by determining their 
structural, mechanical, and optical properties 
\cite{bla98,bra02,mos92,hai01,bra97,bra98,woo02,man01,nun95}
revealing a plethora of ordered (polymerised fullerenes) and disordered (amorphous)
carbon phases (see \cite{bla98} for a review).

Based on electron diffraction experiments, 
shock-compressed fullerite~\cite{hir95},
has been conjectured to be a {\em new form of amorphous diamond}
exhibiting IRO~\cite{bla98}. Furthermore,  
it has been argued that the mechanical properties
are determined by remnants of (partially) intact fullerene  cages
distinguishing these phases from 
tetrahedrally coordinated  amorphous carbon (ta-C) produced by
ion-beam techniques.

However, up to now, conjectures concerning the structural
properties are based on indirect information; 
neither the process of formation  of such structures
nor the nature  of the proposed
intermediate range order is understood microscopically.
Pioneering molecular-dynamics (MD) simulations  have shown that
subjecting fullerite to high pressure and high temperatures
may give rise to a dense amorphous phase~\cite{zha94}.  
However, 
the sample in~\cite{zha94}
had to be compressed to very high densities
($4.4$ g/cm$^3$ at $T=2500$K) corresponding to pressures exceeding $100$GPa,
far beyond any experimentally obtainable conditions. 
Empirically, the transition from fullerite to amorphous carbon
occurs in the range of $10-30$ GPa, depending
on the speed of compression
and on temperature (room temperature in \cite{nun91} and approximately 
$2000$K in
shock experiments~\cite{hir95,bla98}).
How may this contradiction be resolved? There are 
indications that shear stress may play a role 
in the destabilisation of the cages~\cite{nun91}. 
However, at present it is not known which microscopic
processes cause the transition from fullerite
to amorphous carbon.

There is thus a pressing need for a theoretical understanding of
the exact microscopic mechanisms occurring during this phase transition.
To which extent is a  macroscopic  picture of  a rapidly imploding 
spherical shell appropriate?  Is shear important? 
How strong is the dependence of the final structural
properties on the details of the compression process? 
Is there intermediate range order in the final dense carbon phase? 

This letter reports on quantum-mechanical computer simulations 
of shock-compressed
fullerite, starting from fcc-crystallised and 
tetragonally polymerised fullerenes. 
Our findings provide profound insights into the above-mentioned issues
and can be summarised as follows.

{\em Microscopic mechanisms and transition pressure.} Hydrostatic pressures 
transform fullerite into amorphous carbon 
by a succession of three steps: 
{\em (i)} polymerisation mainly by 2+2 ring-formation starting 
at $P\!<\!10$ GPa,
{\em (ii)} breaking of the C$_{60}$ cages starting at $P\!\sim\!40$ GPa,
and {\em (iii)}, at still higher pressures, collapse of the 
broken C$_{60}$ cages via formation of chemical bonds across
the cage. 
Remarkably, process {\em (iii)} occurs very slowly, 
and (c.f. Ref.~\cite{zha94}) 
very high hydrostatic pressures (exceeding $60$~GPa)
are required for a complete collapse.

Non-hydrostatic (external) stresses, by contrast, result in a rapid shear
deformation of the unit cell: steps
{\em (ii)} and {\em (iii)} happen simultaneously and very rapidly.
In other words,
the C$_{60}$ molecules are ground -- 
and not merely crushed to form amorphous carbon.
Application of shear significantly lowers the transition pressure.
We explain this fact in terms of a continuum model (snap-through
instability of a spherical shell under shear).

{\em Microstructure and intermediate-range order.} 
We analyse the resulting
structures after compression to $35$~GPa and subsequent relaxation to 
ambient conditions and find that the morphology 
depends on the amount of applied shear. 
In order to characterise the phase transition and
resulting morphologies it is not sufficient
to merely specify pressure and temperature, shear
is equally important.
At low shear stresses  the mechanical strength of the material is low 
and at the same time, intact C$_{60}$ cages are still present. At high
shear stresses, by contrast, no remnants of C$_{60}$ cages
are observed. In this case,
bulk moduli are found to be large, of the order
of that of diamond. 
The radial distribution functions $G(r)$  indicate that our
fullerite-derived dense phase has lost its nanoscopic order 
and turned into conventional ta-C. 
We show that fine details in $G(r)$ of 
shock compressed fullerite~\cite{hir95} are most likely 
effects of an incomplete experimental 
$k$-space sampling and not a fingerprint of IRO.


\begin{figure}[t]
\psfrag{a}{{\bf a}}
\psfrag{b}{{\bf b}}
\psfrag{y1}[][tc]{$\rho$ [\,g/cm$^3$]}
\psfrag{y2}[][tc]{Number of bonds}
\psfrag{x2}[][c]{$P$ [GPa]}
\includegraphics[width=6cm]{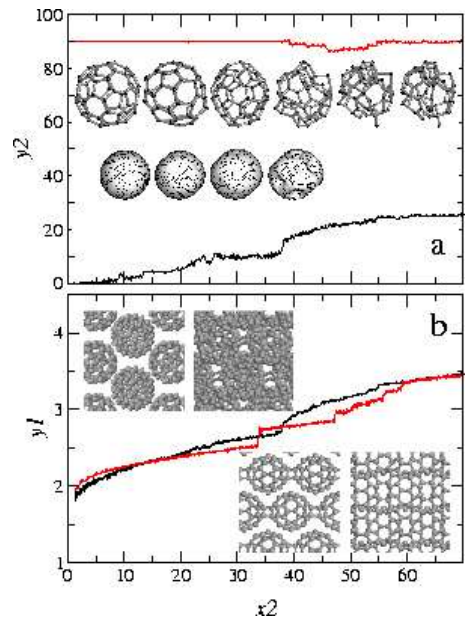}
\caption{\label{fig:1} Fullerite under increasing
  hydrostatic pressure, at $2000$~K.
  {\bf a} Evolution of bonding in the absence of shear stress, starting
  from an fcc crystal of fullerenes. 
   Shown are the number of 
  bonds per C$_{60}$-molecule as a function of applied pressure. 
   Intramolecular (red, upper curve) and intermolecular
  bonds (black, lower curve) are plotted separately. 
  Insets show the shapes of  a representative C$_{60}$ molecule
  as the pressure increases, in steps of $14$~GPa, and the results
  of a continuum model (snap-through instability of
  a spherical shell) at pressures $5, 20, 30$ and $40$
  GPa. The instability occurs between the third and fourth
  configurations.
  {\bf b} Comparison of density evolution of the compressed 
   C$_{60}$ crystal starting from the fcc   
  (black curve) and from a tetragonally 
  polymerised phase (red curve).
  Insets show the initial and  a intermediate configurations 
  (at pressures $0$ and $35$~GPa, respectively)  for the 
  fcc (top row) and tetragonal (bottom row) phase. } 
  \end{figure}

\begin{figure}[t]
\psfrag{a}{{\bf a}}
\psfrag{b}{{\bf b}}
\psfrag{c}{{\bf c}}
\psfrag{y1}[][tc]{$\rho$ [g/cm$^3$]}
\psfrag{x2}[][c]{$P$ [GPa]}
\psfrag{y2}[][c]{Number of bonds}
\psfrag{y3}[][c]{Fraction of sp$^3$-hybridised bonds}
\psfrag{y4}[][c]{$\Delta$ [eV]}
\psfrag{y5}[][c]{$\Delta$ [eV]}
\psfrag{x4}[][c]{$P$ [GPa]}
\psfrag{x5}[][c]{$P$ [GPa]}
\psfrag{x}{\mbox{}\hspace*{-5mm}$P$}
\psfrag{y}{\mbox{}$S$}
\includegraphics[width=6cm]{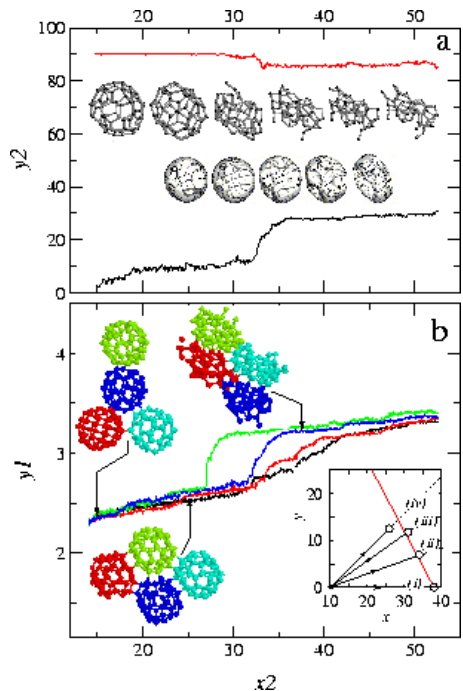}
\caption{\label{fig:2} Fullerite compressed in the presence
of external shear. 
{\bf a} Evolution of the number of bonds per C$_{60}$ molecule (color coding
as in Fig. 1a). Insets displays the morphology of a C$_{60}$-molecule (top)
as the pressure increases, in steps of 6~GPa,
and from a continuum model (bottom).
{\bf b}
Density as a function of pressure in the presence
of external shear $S=\sigma_{xy}$. Shown are four
plots corresponding to four pathways in the 
$S$-$P$-plane as given in the inset: {\em (i)} black,
{\em (ii)} red, {\em (iii)} blue, {\em (iv)} green.
Locations of the collapse transition are shown
as points in the $S$-$P$-plane. Also shown
is the critical line (red) as determined
from a continuum model.
Finally, morphologies of the C-atom configurations in a unit cell for 
case {\em (iii)} are also shown.
} 
\end{figure}

\begin{figure}[htb]
\psfrag{e}{{\bf e}}
\psfrag{a}{{\bf a}}
\psfrag{b}{{\bf b}}
\psfrag{c}{{\bf c}}
\psfrag{d}{{\bf d}}
\psfrag{i}{{\bf i}}
\psfrag{ii}{{\bf ii}}
\psfrag{iii}{{\bf iii}}
\psfrag{iv}{{\bf iv}}
\psfrag{v}{{\bf v}}
\psfrag{vi}{{\bf vi}}
\psfrag{y1}[][c]{$G(r)$ [\AA{}$^{-2}$]}
\psfrag{x1}[][c]{$r$ [\AA{}]}
\includegraphics[width=6cm]{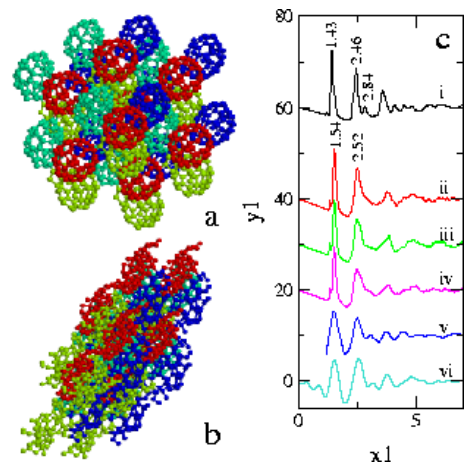}
\caption{\label{fig:3} Structures of fullerite relaxed from
 from $P$$=$35~GPa,$T=2000$~K to ambient conditions.
  {\bf a-b} morphologies of the fcc unit cells 
  resulting from compression with two different
  shear values: (a) no shear and 
  (b) shear according to path {\em (iv)} in the S-P-plane displayed in Fig.~2b.
  {\bf c} radial distribution function $G(r)$ averaged over 20 ps  along
   a trajectory at ambient conditions: 
   ({\bf i}) sample after a hydrostatic compression to $P=35$ GPa, 
   ({\bf ii}) same compression but with high shear,
   ({\bf iii})  reference sample  without any intermediate range order
     (see text),
   ({\bf iv}) after hydrostatic compression to $P=60$ GPa,  
   ({\bf v}) experimental G(r) of shock-treated fullerite~\cite{hir95},
  ({\bf vi}) same curve as in {\bf ii}, convolved with a
  peak-shape function $\sin(Q r)/(\pi r)$ with
  $Q=12.0$\AA{}$^{-1}$.}
\end{figure}

{\em Methods}.
We simulate
the MD of shock-compressed fullerite within an fcc and a tetragonal unit cell 
containing four and two C$_{60}$, respectively.  
Interatomic forces are determined within a non-orthogonal density-functional-based 
tight-binding scheme~\cite{por95}. 
A Langevin thermostat is used to keep the temperature 
close to the $2000$~K
reported for shock conditions~\cite{bla98}. 
Spontaneous shearing of
the simulation cell is allowed for and external  shear stresses 
were applied by using a Rahman-Parrinello~\cite{ram80} barostat. 
After thermalisation of the sample, the external pressure is increased
linearly from $0$ to $70$~GPa within $100$~ps. 
Bond formation and breaking as well as sp$^3$/sp$^2$ hybridisation 
is determined based on 
a population analysis of the bond orbitals, similar to the 
Mulliken charge analysis~\cite{els98}. 
In the remainder of this letter we explain how the results summarised
above follow from our simulations
(Figs.~\ref{fig:1} to \ref{fig:3}).

{\em Hydrostatic conditions.}
Fullerite proves to be very stable under
isotropic compression, withstanding high pressures. 
The microscopic mechanism of collapse is      apparent from 
Fig.~\ref{fig:1}a displaying the evolution of inter- and intramolecular
bonds and the representative fate of one of the cages. 
Starting from an fcc-ordered crystal,
the structure changes gradually through
polymerisation of C$_{60}$-molecules 
(see the rise in the number intermolecular  bonds for $P>5$~GPa, 
lower curve of Fig.~1a).  Double bonds in neighbouring cages break and 
rehybridise with each other to form
a mainly four-membered rings (2+2 cycloaddition~\cite{rao93,polynote}). 
 This process is completed for
 $P\!>\!26$~GPa and the number of intermolecular bonds stays 
 constant up to $P_{\rm c} = 38$~GPa (Fig. 1a).
 
 Upon increasing the pressure beyond 
 $P_{\rm c} =  38$~GPa, the cages break (inset in Fig.~1a), indicated by a drop
 in the number of intramolecular bonds (upper curve in Fig.~\ref{fig:1}a)
  and form new bonds to neighbouring fullerenes (sudden raise
 of the lower curve in  Fig.~\ref{fig:1}a) leading to a densification 
 of the material (Fig.~1b). 
 At even higher pressures, the C$_{60}$ cages gradually collapse forming bonds 
 inside the cages marked by a steady increase
 in the number of intramolecular bonds in Fig.~\ref{fig:1}a 
 for large values of $P$.
 When and how fast this process occurs depends on the local 
 alignments of the cages. 

In order to study the influence of the initial configuration 
on the phase transition, a second simulation 
 starting from a tetragonally polymerised state (lower inset in Fig. 1b) 
 has been performed.
The resulting density-pressure curve (Fig.~1b) is qualitatively unchanged at
high pressures, as is the resulting structure  (at pressures exceeding
$60$~GPa). In the intermediate regime, however, the structures are quite
different (insets of Fig.~\ref{fig:1}b), as are the density-pressure
curves (Fig.~\ref{fig:1}b): in the tetragonal phase, the C$_{60}$
molecules cannot rotate freely (as opposed to those in fcc phase).  
Intermediate configurations are thus seen to be very regular,
with a high degree of polymerisation.

Regardless of the initial conditions,  unrealistically high pressures
(exceeding $60$~GPa) are required
to entirely destroy the cages (resulting in a dense amorphous phase after
relaxation to ambient conditions). This is not consistent
with the experimentally found transition pressures.

  
{\em Non-hydrostatic conditions.} 
We have compressed the polymerised fcc phase
under non-hydrostatic conditions, along 
three paths - {\em (ii)}, {\em (iii)}, and {\em (iv)} - 
in the $S$-$P$ plane shown in the inset of Fig.~2b ($S = \sigma_{xy}$ is the shear stress
in the (001) plane).  Path {\em (i)} corresponds to the hydrostatic
case.
Fig.~2a shows the  numbers of 
intra- and intermolecular bonds per C$_{60}$ molecule, for  case {\em (iii)}.
Comparison with Fig.~1a and inspection of the pressure-density
relationship in
Fig.~2b  confirms that shear plays an important role lowering the critical
pressure and increasing the speed of the phase 
transition which happens now within a few GPa
at approximately $34$, $31$, and $26$~GPa for 
cases {\em (ii)}, {\em (iii)}, and {\em (iv)}. 
The critical shear values
at the collapse are $S_{\rm c} = 6.94, 11.8$ and $12.5$
and shown as points in the $S$-$P$-plane in Fig.~2b.

Fig.~2a also indicates
that the microscopic mechanism is different from the
zero-shear case described above: in the presence of shear
the cages are ground (see the transition from a spherical to an oblate
cluster in the inset of Fig. 2a). After a rapid transition the number
of intra- and intermolecular bonds remains 
constant (as opposed to 
Fig.~\ref{fig:1}a).
The cages are so strongly flattened that the number of
intramolecular bonds cannot be increased by raising the pressure further.

The final structure however (at pressures $>35$~GPa) is
similar to the structure obtained under hydrostatic
conditions at pressures exceeding $60$~GPa.

\begin{table}
  \begin{tabular}{cccc}
    \hline\\[-3mm]\hline\\[-2mm]
    \hspace*{2mm}Load mode\hspace*{2mm}
        &\hspace*{2mm} $\rho$ [g/cm$^3$] \hspace*{2mm} & B [GPa]   
	& \% sp$^3$ \\[2mm]
    \hline\\[-2mm]
    (i) & 2.1 & 50 & 18\%  \\
    (ii) & 2.3 & 60 & 20\%  \\
    (iii) & 3.0 & 210 & 69\%  \\
    (iv) & 3.1 & 320 & 79\%  \\
    \hline\\[-3mm]\hline
  \end{tabular}
  \caption{\label{tab:1}
  Characterisation of fullerite phases after compression to 35 GPa  
  and relaxation to ambient conditions. Load conditions as in
  Fig.~\ref{fig:2}b.
} 
\end{table}


{\em A continuum model.}
To which extent can the phase transition in Figs.~1 and
2 be understood in terms of a continuum model?
   We have modeled a $C_{60}$-molecule in fullerite as a spherical shell
     with diameter $0.71$nm. Its Young modulus ($5.32$ TPa)
      and effective thickness ($0.08$nm) are determined from
      tight-binding simulations of a graphene sheet. Forces 
        are applied at $12$ nodes corresponding         to the fcc
        symmetry of the fullerite crystal.  The shell 
          exhibits
          a snap-through instability at $P_{\rm c} = 37.8$ GPa  
	  (see snapshots of the shell in Fig. 1a)   
          to be compared with $38$ GPa obtained above. 
          It must be noted that the agreement is better than
            expected \cite{note}, and that the continuum model
             cannot explain the dynamics and morphologies at
              pressures beyond the collapse (see below).

The decrease of the critical pressure
with increasing shear stress can also be understood in terms
of the continuum model (inset of Fig.~2a): shear
destabilises a spherical shell under pressure and
lowers the critical pressure. The continuum model
predicts that critical
shear stress and pressure are linearly  related,  the
result is shown as a red line in the inset of Fig.~2b,
in good agreement with the atomistic results
for not too large values of shear stress.
This shows that the onset of the instability
 of fullerite under shear stress can be described 
 by elasticity theory. The ensuing collapse, however,
 must be modeled microscopically (in terms of breaking
 of bonds and rehybridisation).
 Finally, the continuum model fails at high shear stresses
 (c.f. path {\em (iv)} in the inset of Fig.~2b).

{\em Resulting morphologies.} The question remains:
what structures are obtained 
when the compressed fullerites are relaxed from -- for instance -- $35$ GPa and $2000$~K 
to ambient conditions? 
The  morphology of the relaxed phase 
depends clearly on the amount of applied shear. 
Consider the relaxed structures shown in Fig.~\ref{fig:3}a and b.
They were obtained from samples compressed along the S-P-path {\em (i)} and {\em (iv)} (as shown 
in the inset of Fig. 2b), respectively. 
For no or low shear stress, densities, bulk moduli and sp$^3$ fractions  are small (see load mode {\em (i)}
and {\em (ii)} in Tab.~1),
polymerisation is evident but the cages are
still intact (Fig.~\ref{fig:3}a). At high shear (load mode {\em (iii)}
and {\em (iv)} in Tab.~1), 
by contrast, densities, bulk moduli and sp$^3$ fractions  are much larger and 
no remnants of C$_{60}$ cages are 
found (Fig.~\ref{fig:3}b).

Does the  radial distribution function
reflect possible IRO in the relaxed samples?
Curves {\bf i} and {\bf ii} in Fig.~\ref{fig:3}c display $G(r)$ for 
the two cases discussed above.
In the zero-shear case, C$_{60}$-specific 
distances are still present (
the
 average nearest-neighbour bond length in graphite is $0.142$~nm) 
thus showing clear evidence of IRO.
 In the other sample, by contrast, the 
radial distribution function is clearly different exhibiting
 the first peak  at $0.154$~nm, as in diamond.
Curve {\bf iii} in Fig. 3c corresponds to a 
sample relaxed from $P=60$ GPa, $S=0$ GPa
 to ambient conditions. Obviously this phase has a structure similar
 to {\bf ii}.
  
To test the hypothesis of a  {\em new form of amorphous
diamond}, we have created reference sample exhibiting no IRO, 
obtained by isochorus heating of $240$ carbon atoms to $10000$~K
 while monitoring the system for sufficient chaotic motion
 followed by a  relaxation to ambient conditions. It exhibits 
 ta-C characteristics and approximately the 
same density and sp$^3$/sp$^2$ hybridisation as the densest structure 
in Tab.~1
Surprisingly, the resulting $G(r)$ (curve {\bf iv} in Fig.~\ref{fig:3}c) is
essentially identical to {\bf ii} and consequently, the radial
distribution function {\bf ii} in Fig.~3b provides no
fingerprint of IRO.

The conditions of the simulations reported here are
close to those in the shock-compression experiments
reported in ref.~\cite{hir95}.
The third peak in the  empirically determined $ G(r)$ (curve {\bf v} in 
Fig.~\ref{fig:3}c) at approximately $0.312$~nm
has been accredited to the diameter of six-membered rings in the
C$_{60}$ molecules or to a slightly elongated third-neighbor
distance for diamond. However, it should be kept in mind
that experimental radial-distribution functions are 
generated from reciprocal space scattering
information which is limited by an inevitable momentum cut-off Q.
Thus, for the sake of comparability, the  simulated $G(r)$ 
of our dense amorphous carbon
(curve {\bf ii}) 
was convoluted with an appropriate cut-off function reflecting the experimental
$Q$$=$$12$ \AA{}$^{-1}$~\cite{hir95}. 
The resulting radial distribution function (curve {\bf vi} in Fig. 3c) 
is very similar to that
observed in~\cite{hir95}.
We conclude that the third peak in {\bf v} is unlikely to reflect IRO,
it is likely to be an effect of the finite experimental resolution.

Support from the Fraunhofer MAVO MMM-Tools 
and Vetenskapsr\aa{}det is gratefully acknowleged.
We thank T. Frauenheim for his Hydrocarbon-Slater-Koster tables.
Computations were 
performed on the CEMI cluster of the Freiburg Fraunhofer
Institutes EMI/ISE/IWM.


\begin{thebibliography}{23}
\bibitem[{not()}]{note}
\bibinfo{note}{A C$_{60}$-cage consists of $5$- and $6$-rings whereas the
  elastic constants were taken to be those of a graphene sheet (consisting of
  $6$-rings only).}
\end{thebibliography}
\end{document}